\documentclass[9pt, conference]{IEEEtran}
\IEEEoverridecommandlockouts

\usepackage{cite}
\usepackage{amsmath,amssymb,amsfonts}
\usepackage{algorithmic}
\usepackage{graphicx}
\usepackage{textcomp}
\usepackage{xcolor}
\usepackage{hyperref}
\usepackage{booktabs}
\def\BibTeX{{\rm B\kern-.05em{\sc i\kern-.025em b}\kern-.08em
    T\kern-.1667em\lower.7ex\hbox{E}\kern-.125emX}}
\usepackage{threeparttable}

\begin{document}

\title{Learning Music Audio Representations With Limited Data\\
  \thanks{Christos Plachouras is supported by UK Research and Innovation
      [grant number EP/S022694/1].}
}
\author{\IEEEauthorblockN{Christos Plachouras, Emmanouil Benetos, Johan Pauwels}
\IEEEauthorblockA{\textit{School of Electronic Engineering and Computer Science}\\
\textit{Queen Mary University of London}\\
London, United Kingdom\\
\{c.plachouras, emmanouil.benetos, j.pauwels\}@qmul.ac.uk}}

\maketitle
\begin{abstract}
  Large deep-learning models for music, including those focused on learning general-purpose music audio representations, are often assumed to require substantial training data to achieve high performance. If true, this would pose challenges in scenarios where audio data or annotations are scarce, such as for underrepresented music traditions, non-popular genres, and personalized music creation and listening. Understanding how these models behave in limited-data scenarios could be crucial for developing techniques to tackle them.

  In this work, we investigate the behavior of several music audio representation models under limited-data learning regimes. We consider music models with various architectures, training paradigms, and input durations, and train them on data collections ranging from 5 to 8,000 minutes long. We evaluate the learned representations on various music information retrieval tasks and analyze their robustness to noise. We show that, under certain conditions, representations from limited-data and even random models perform comparably to ones from large-dataset models, though handcrafted features outperform all learned representations in some tasks.
\end{abstract}

\section{Introduction}
Large models tackling audio-based Music Information Retrieval (MIR) tasks are commonly thought to require substantial amounts of training data \cite{mcfeeSoftwareFrameworkMusical2015, ponsTrainingNeuralAudio2019}.
This presents a challenge, as obtaining music data and annotations often requires human expertise and extensive effort. The issue is amplified for less popular or underrepresented music traditions and styles for which few music recordings and little music-theoretic knowledge to annotate them might be available. The problem is also present in personalization, where information about listening or music-making preferences for a music streaming service user or an artist respectively might be scarce \cite{ferraroMusicColdstartLongtail2019}.

A category of large music models that are specifically considered data-hungry is representation learning models. These aim to learn compact and informative representations of the input data that are useful for a variety of so-called downstream tasks. Representation learning for music audio has been a popular approach in MIR in recent years particularly because, for the downstream tasks, the large representation model is used as a feature extractor, allowing a computationally efficient transfer of knowledge. Various architectures and training paradigms have been investigated, sometimes with music-specific modifications. Representation learning presents some unique considerations: representations are not always evaluated on their training objective, and the downstream model utilizing the representation can affect how much relevant information for the task can be extracted from the representation.

In spite of the prevalence of music audio representation learning models, little is known about their behavior in scenarios where few data are available for training, and their supposed data requirement has not yet been demonstrated with experimental evidence in the music domain. The impact of the training paradigm and the model architecture on the model's ability to effectively learn good representations is also not well understood in practice. Additionally, some other considerations about representation learning arise through the context of limited data. For example, how important is it to learn a good representation in the first place? Can a downstream model ``recover'' performance from a bad representation?

In this work, we conduct the first systematic evaluation of music audio representation learning models trained with limited data. We train 5 popular music representation models that differ in architecture, training paradigm, input length, and input representation on music collections ranging from 5 minutes to $\approx$8000 minutes long. We use them, along with untrained models, as feature extractors in 3 downstream MIR tasks: music tagging, instrument recognition, and key detection. We evaluate their performance, their robustness to noise, and their `linear separability' for a given task, loosely defined as how successfully a linear classifier can separate the feature space into the relevant classes. Our findings indicate that performant representations can often be learned from just a few minutes of audio using existing models. However, they also raise concerns about the quality and real-world suitability of learned representations, given the significant impact of the downstream setup on downstream performance.

\section{Related work}

While the training data requirements for audio representation learning models have not been systematically reviewed, Pons et al. have investigated the performance of end-to-end audio classifiers trained with different amounts of data \cite{ponsTrainingNeuralAudio2019}. They use existing classifiers with dropout and L2 regularization and compare them to prototypical networks and transfer learning approaches. They find that conventional end-to-end models are usually outperformed by prototypical networks and models benefiting from transfer learning, although this is sometimes not the case when there is enough data available ($>$100 examples per class). In a setting similar to representation learning \cite{ponsRandomlyWeightedCNNs2019}, they extract representations from untrained convolutional networks, which they use to train shallow classifiers on audio auto-tagging. They demonstrate that these systems achieve good performance, comparable to handcrafted features and trained models, and suggest that the model's structure itself contributes significantly to an end-to-end model's performance.

While we evaluate existing music representation learning models on the limited-data use case, it is worth noting that various approaches have been developed to learn from small audio datasets. Data augmentations that preserve the semantic validity of the audio have been used to increase the number of training samples and improve performance in sound event detection \cite{salamonDeepConvolutionalNeural2017, piczakEnvironmentalSoundClassification2015} and acoustic scene classification \cite{munGenerativeAdversarialNetwork2017}. Approaches leveraging cross-domain or cross-dataset transfer learning have also been explored, such as for optical music character recognition with a network pretrained on handwriting \cite{paulEnsembleDeepTransfer2022}, music emotion recognition with one pretrained on speech \cite{gomezcanonTransferLearningSpeech2020}, and music tagging with pretraining on different music traditions \cite{papaioannouWestEastWho2023}. Finally, many approaches have investigated incremental learning in cases where few examples of new classes are available, such as for few-shot instrument recognition \cite{garciaLeveragingHierarchicalStructures2021} and source separation \cite{wangFewShotMusicalSource2022}.




\section{Experiment design}
\begin{figure}
    \centering
    \includegraphics[width=0.95\linewidth]{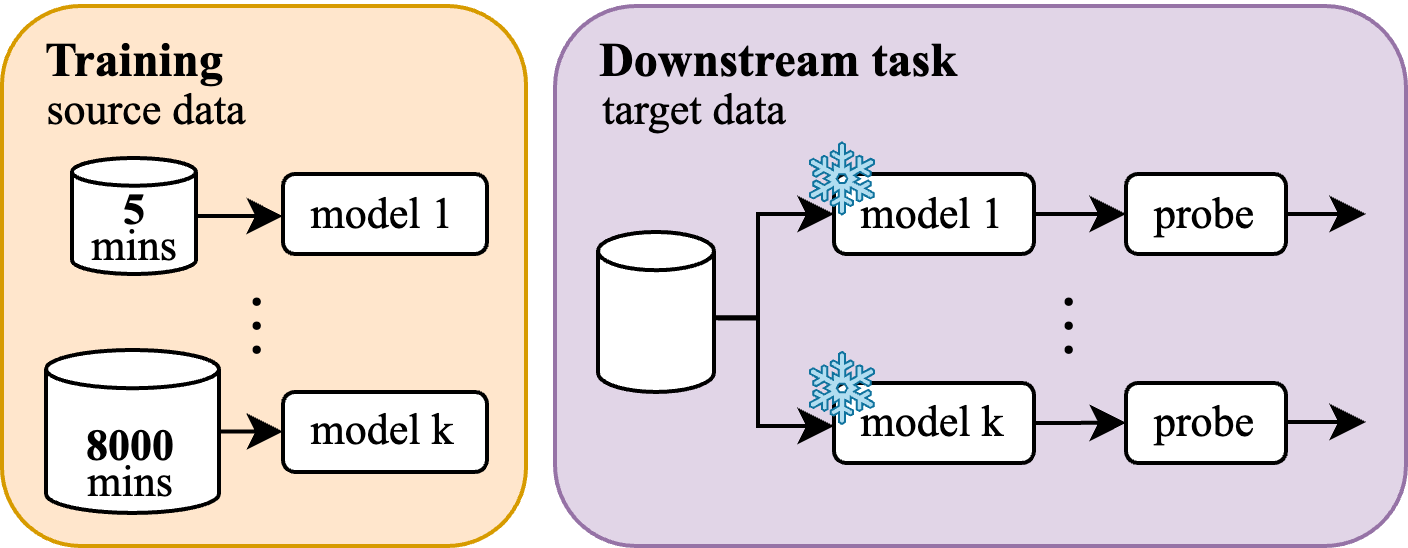}
    \caption{Flow diagram of representation learning with limited data and evaluation on downstream tasks.}
    \label{fig:flow}
\end{figure}

An overview of the key experiment components is presented in Fig. \ref{fig:flow}. Each model is trained on a number of dataset subsets. They are then used as feature extractors, meaning their parameters are frozen, coupled with shallow classifiers (probes) to solve downstream MIR tasks, including in perturbed data conditions. All relevant code to reproduce this study is openly available.\footnote{Code repository: \url{https://github.com/chrispla/limited-music-representations}}

\subsection{Models}
We employ a range of audio representation learning models that have already been used in the music domain, that differ in architecture, training paradigm, input representation, and input length. Given the difficulty of training thousands of model parameter combinations to disentangle all performance factors, we opted to prioritize model configurations from the original papers when these exist, following the authors' original design intentions. This approach aligns with how other researchers and engineers are likely to use these models in their real-world applications and data: making few or no modifications to the architecture. A summary of key configuration aspects is presented in Table \ref{table:models}.

\begin{table*}
\caption{Model and Training Information}
  \centering
  \begin{tabular}{lcccccccccc}
    \toprule
    & \multicolumn{3}{c}{\textbf{Model}} & \multicolumn{3}{c}{\textbf{Input}} & \multicolumn{4}{c}{\textbf{Training}} \\
    \cmidrule(lr){2-4} \cmidrule(lr){5-7} \cmidrule(lr){8-11}
    \textbf{Name} & \textbf{Architecture} & \textbf{Param.} & \textbf{d$_{rep.}$} & \textbf{Length} & \textbf{Feature} & \textbf{Mel bins} & \textbf{Paradigm} & \textbf{Task} & \textbf{\textit{lr}} & \textbf{\textit{wd}} \\
    \midrule
    VGG \cite{simonyanVeryDeepConvolutional2015} & CNN & 3.7m & 512 & 3.75s & mel spec. & 128 & Supervised & Tagging & 1e-3 & 1e-5 \\
    MusiCNN \cite{ponsMusiCNNPretrainedConvolutional2019} & CNN & 12.0m & 200 & 3.00s & mel spec. & 96 & Supervised & Tagging & 1e-3 & 1e-5 \\
    AST \cite{gongASTAudioSpectrogram2021} & Transformer & 87.0m & 768 & 5.12s & mel spec. & 128 & Supervised & Tagging & 1e-4 & 1e-5 \\
    CLMR \cite{SSL_Spijkervet_2021} & CNN & 2.5m & 512 & 2.68s & waveform & - & Self-Supervised & Contrastive & 3e-4 & 1e-6 \\
    TMAE \cite{jangModalityAgnosticSelfSupervisedLearning2023a} & Transformer & 7.2m & 256 & 4.85s & mel spec. & 96 & Self-Supervised & Masked Modeling & 1e-4 & 1e-4 \\
    \bottomrule
  \end{tabular}
  \begin{tablenotes}
    \small
        \item \hspace{1.2em} \textbf{d$_{rep.}$}: dimensionality of the layer representation chosen, \textbf{\textit{lr}}: learning rate, \textbf{\textit{wd}}: weight decay
    \end{tablenotes}
  \label{table:models}
\end{table*}

We train 3 models in a supervised manner on music auto-tagging: VGG, MusiCNN, and AST. VGG \cite{simonyanVeryDeepConvolutional2015} is a convolutional model used in multiple domains, including audio \cite{ponsMusiCNNPretrainedConvolutional2019}. We follow an existing VGG-like implementation \cite{Won_2020}, using 7 2D convolutional layers with an increasing number of 3$\times$3 kernels and using their final output as the feature layer. MusiCNN \cite{ponsMusiCNNPretrainedConvolutional2019} instead utilizes carefully designed, musically-motivated kernels and pooling to capture timbral and temporal information and be pitch invariant. We use the original implementation, including the recommended feature layer after pooling. Finally, the Audio Spectrogram Transformer (AST) \cite{gongASTAudioSpectrogram2021} is an adaptation of the Vision Transformer (ViT) \cite{dosovitskiyImageWorth16x162021} to audio operating on the mel spectrogram. We implement AST similarly to an adaptation of it to music \cite{papaioannouWestEastWho2023}, using 12 transformer blocks and a 768-dimensional representation and the original patch embeddings proposed.

We also train 2 self-supervised models: CLMR and TMAE. CLMR \cite{SSL_Spijkervet_2021} is an adaptation of the SimCLR contrastive learning approach \cite{SimCLR_2020} to self-supervised music audio representation learning. It uses an existing convolutional encoder \cite{tag_wave_0} that operates on raw audio waveforms, consisting of 9 1D convolutional layers with a kernel size of 3, and learns representations by building invariance to musically relevant data augmentations. We also implement a masked autoencoder with a ViT-based encoder and decoder, which we refer to as TMAE, similar to an existing modality-agnostic baseline model \cite{jangModalityAgnosticSelfSupervisedLearning2023a}. We use 12 encoder layers with 256-dimensional representations and 6 decoder layers with 128-dimensional representations, a setup we found to work well in non-data-limited scenarios.

\subsection{Pre-training}
\subsubsection{Data}
To train the models, we create subsets of the MagnaTagATune (MTAT) dataset \cite{magnatagatune}. We decided to simulate limited-data scenarios on MTAT rather than using smaller datasets since it is a dataset that has been widely used in MIR, helping put our results into perspective, it has enough tracks to allow us to assess different levels of data constraints, and it is easily accessible, unlike many smaller datasets. It contains 25,863 music clips of around 30 seconds, each annotated with at least one tag. The tags vary in category, ranging from genres and moods to instruments and other musical concepts. When used for music auto-tagging, only the 50 most popular tags are usually considered, but clips not containing any of these tags are also included in the dataset.

To create subsets for limited-data training, we first create a set $T$ containing 10 tags from the 50 most popular ones that loosely represent the most prevalent genres. Then, we randomly sample $n$ clips per tag $t \in T$, where $n \in \{1, 2, 5, 10, 20, 50, 100\}$, following Pons et al. \cite{ponsTrainingNeuralAudio2019}. This means, for example, that for $n=1$ there are a total of 10 clips, 1 from each genre, with a total subset duration of 5 minutes. For $n=100$, there are 1000 clips forming a subset that is 500 minutes, or around 8 hours long. Notably, following the finding that the model's architecture itself might considerably contribute to the representation's content \cite{ponsRandomlyWeightedCNNs2019}, we also consider the case in which the models are initialized but not pretrained, which we denote as $n=0$. The subset selection is done solely from the training splits used for music auto-tagging so that there is no overlap during the downstream task evaluation. We use the most common MTAT splits \cite{Won_2020}, which ensure clips from the same track can only belong to the same split. Finally, for comparison reasons, we also train the models on the entire training split of MTAT, which we indicate as \textit{full}.



\subsubsection{Training and stopping}
Similar to existing work \cite{ponsTrainingNeuralAudio2019}, we opt not to use a validation set during pretraining. The primary reason for this is that in such data-constrained scenarios, not enough data to create a useful validation set is always available. Further, since we are dealing with representation learning models with different training paradigms, the validation set would validate different objectives for each model. These objectives are, in most cases, not aligned with those of the downstream tasks, for which we do use a validation set. The models are significantly larger than those tested in existing work \cite{ponsTrainingNeuralAudio2019}, and are thus likely to overfit in the very data-limited scenarios. However, that allows us to stick to original implementations (including regularization configurations) to test how much performance we can recover in the downstream task out of ``non-ideal'' data representations. We train each model with the Adam optimizer (see Table \ref{table:models} for more information) until the training loss has not improved for 100 epochs. For the batch size, we use the largest power of 2 which allows the model to fit in a consumer graphics card with 8 GB of memory.



\subsubsection{Baselines}
To give context to the performance of the models, we also consider two popular baselines of handcrafted features: Mel-frequency cepstral coefficients (MFCC) and Chromagrams. MFCCs are derived from the discrete cosine transform of the log mel spectrogram, capturing the overall shape of the spectrum. They have been widely used in MIR for their ability to capture timbral information while having some built-in pitch invariance. We use 13 coefficients along with their first and second derivatives and take their clip-level average, forming a 39-dimensional feature vector. Chromagrams express the salience of each pitch class in a clip. They are used for their ability to capture harmonic information while being robust to timbre changes. For simplicity, we use 12 pitch classes, assuming most clips are tuned in a 12-tone equal temperament system, but this can easily be adjusted for datasets with clips using a different tuning system. We take their clip-level average, forming a 12-dimensional feature vector.

\subsection{Downstream setup}
The most common way to evaluate music representations is to use them as features for a variety of downstream MIR tasks, which directly addresses how useful the representations are for these tasks. Further, lower-level tasks such as monophonic instrument recognition and pitch estimation can provide insights into the music information that is encoded in the representations. To implement the downstream tasks, we use the \texttt{mir\_ref} library \cite{plachourasMir_refRepresentationEvaluation2023} that allows us to conduct the thousands of downstream combinations in a reproducible way, and \texttt{mir\_eval} \cite{mir_eval} that provides standard metric implementations. We run each downstream training 3 times and report the mean result for each metric.

\subsubsection{Tasks}
We select three common MIR tasks requiring different types of music information. We use MagnaTagATune in a downstream music auto-tagging task, using the same train-test splits to ensure test tracks are absent from both training and downstream training. For the music tagging models, this setting resembles end-to-end music tagging. However, in practice, the models are only trained on small subsets of the dataset, and only their representations are used for the downstream task, so the generalizability of the representation is still tested. We use the area under the receiver operating characteristic curve (AUC-ROC), the area under the precision-recall curve (AUC-PR), and the per-class accuracy as evaluation metrics, although we visualize the most common of these (AUC-ROC) due to space constraints and all metrics showing very similar trends.\footnote{Full results are available online in the associated code repository.} To test the representation's encoding of timbral information at a lower level, we use the TinySOL dataset \cite{orchideasol} for monophonic instrument recognition, which contains 2,913 recordings of single-note strokes from 14 instruments. Finally, to test the representation's encoding of pitch information, we use the Beatport dataset \cite{beatport, beatport_phd} for global key detection. We discard clips that contain multiple or no annotations, ending up with 1,272 clips. For both datasets, we use a fixed-seed 80-10-10 stratified train-validation-test split. We report the F1 score for TinySOL, and a commonly-used, musically-motivated weighted score for Beatport \cite{mir_eval}. 

\subsubsection{Downstream models}
We decide to use two models: a single-layer perceptron (SLP), and a multi-layer perceptron (MLP) with a hidden layer with 256 units followed by one with 128. This choice follows existing experimental evidence that the downstream model capacity can influence performance, and that 2-hidden-layer MLPs can extract more performance out of a representation than 1-hidden-layer MLPs and SLPs \cite{plachourasMir_refRepresentationEvaluation2023}. Both models are trained with the Adam optimizer, a learning rate of 5e-3, and 200 epochs, with the best model picked using the validation set. For both training and evaluation, we average the representations on a clip level.

\subsubsection{Robustness}
Understanding the inherent robustness of representations to common audio perturbations is critical for real-world system deployments, but it is often overlooked in MIR. For each dataset, we create perturbed versions with Additive Gaussian White Noise (AGWN) at a signal-to-noise ratio of 0 dB. While this results in very noisy audio, it's a level of degradation that would likely not affect a human's performance on these tasks and is sometimes encountered in archival recordings and real-world music identification. We only use these versions for evaluation and not for probe training, in order to test the representations' inherent robustness, rather than how well the probes can build invariance to them.

\section{Results}
\begin{figure*}
  \centering
  \includegraphics[width=0.94\textwidth]{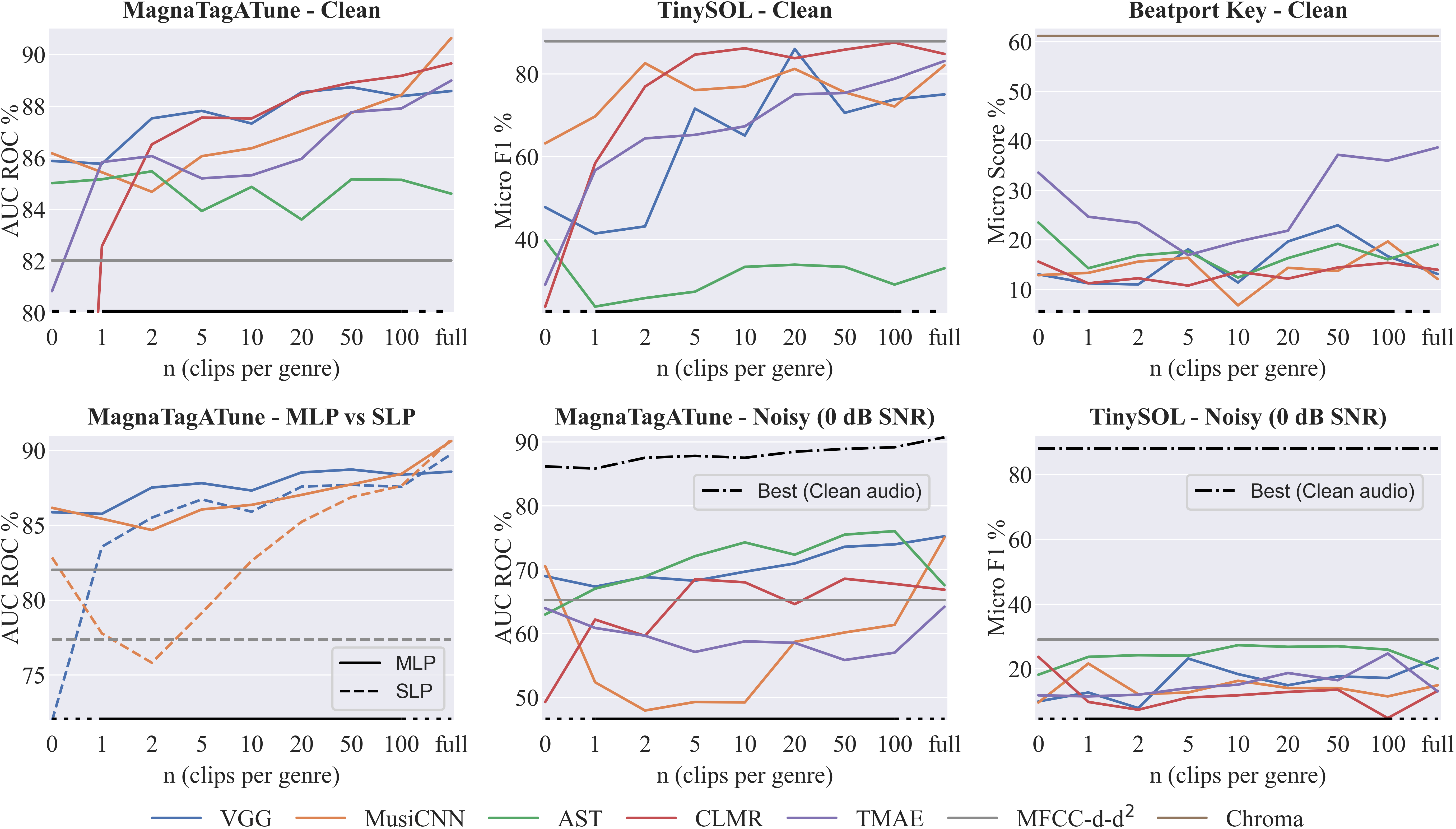}
  \caption{Downstream performance of all features. The first row contains results using the original audio and the MLP probe. The second row, from left to right, contains a comparison of performance between probes, and performance on MTAT and TinySOL with noisy audio using the MLP probe.}
  \label{fig:mega}
\end{figure*}
We present the results for different representations, downstream tasks, and the noise perturbation in Fig. \ref{fig:mega}. In all plots, the x-axis is roughly in geometric scale, representing subsets from 5 minutes ($n=1$) to 500 minutes ($n=100$) long, with the fully-trained (\textit{full}) and random ($n=0$) models also included.

The first row contains results on the original datasets, using the MLP probe. On MagnaTagATune, we observe that, even with just 5 minutes of training data, almost all representations are performing reasonably well. Specifically, there is a 0.02 AUC ROC average difference between $n=1$ and $n=100$ across all representations. Similarly, the difference between $n=100$ (500 minutes) and $n=$ \textit{full} ($\approx$8,000 minutes) is less than 0.02 AUC ROC on average. AST sees no performance improvement with more data and is the worst-performing model for $n\geq5$. This is consistent with previous findings that AST needs a lot more training data, possibly through pretraining in other domains, to reach its full potential \cite{gongASTAudioSpectrogram2021, chenEmpiricalStudyTraining2021}. TMAE, the other transformer tested, however, consistently outperforms AST at almost every data subset, and is the best-performing representation for 5 minutes of training data ($n=1$). Given that AST likely benefits in this task from being trained on music tagging, this result might be attributable to TMAE's much smaller size (7.2m vs 87m parameters), rather than its training paradigm. CLMR performs well when enough data are available, but suffers notably for $b\leq2$.

A notable result is the performance of representations extracted from non-trained models. Using a 2-hidden-layer MLP, representations from MusiCNN and VGG reach around 0.86 AUC ROC and 0.35 AUC PR. This setup resembles principles from reservoir computing \cite{zhangSurveyReservoirComputing2023, shenReservoirTransformers2021}, in which a signal is mapped into a high-dimensional space, and a system is trained to map it to a desired output. It also further strengthens existing results in randomly initialized CNNs for audio classification \cite{ponsRandomlyWeightedCNNs2019}, and raises the question of how ideal a representation needs to be when the downstream seems to be able to recover relevant information. 

Similar behaviors are observed for instrument classification. On average the performance difference with more data is relatively small, but the individual behaviors vary. CLMR, again, starts obtaining good results after being trained with more than 25 minutes of data, with TMAE consistently improving with more data but only reaching performance close to CLMR when fully trained. The key detection task showcases the representations' poor pitch recognition, as they all perform significantly worse than the chromagram baseline. The amount of training data used, overall, seems to have little impact on performance. CLMR and MusiCNN, the design of which targets pitch invariance, are in fact the worst performers.

On the left of the second row, we see the performance from the VGG, MusiCNN, and MFCC features using the two different probes tested. We observe a huge performance difference, particularly for MusiCNN, in the zero and very limited data cases. In fact, the SLP MusiCNN performance initially decreases from the random model, potentially indicating that MusiCNN overfitted severely in the very limited data cases. However, the MLP's performance remains relatively stable, in spite of the overfitting. This further supports the theory that a capable enough downstream model might be able to recover relevant information from a bad representation. The other representations presented much smaller differences between probes (\textless 0.03 AUC ROC on average) and were therefore omitted for clarity. Very similar behavior for all representations was also present in TinySOL, although was less prevalent for Beatport.

The introduction of white noise to the audio significantly deteriorates performance across all models. AST demonstrates the highest resilience, while the overfitted MusiCNN experiences the most severe decline, with even the MLP unable to recover its performance. The impact of training data volume follows similar trends for each model as observed in the clean audio scenario, albeit with more pronounced effects. Notably, the performance degradation persists even with full dataset training, underscoring the need for further research and improvement in model robustness for real-world applications.

\section{Conclusions}
Our analysis shows that even random models can seemingly produce performant representations. However, the choice of downstream model (SLP vs MLP) had a notable impact on performance, particularly in very limited data scenarios. This suggests that a capable downstream model might be able to compensate for limitations in the learned representations to some extent, challenging what a ``good'' representation is. Further, for any dataset size, representations were generally not robust to white noise, which raises questions about the relevance of the information they are encoding and their suitability for real-world applications. Finally, in monophonic instrument recognition and global key detection, low-dimensional, handcrafted features were matching or even outperforming all representations.

Further experiments are needed to disentangle the impact of the model architecture, the training paradigm, regularization techniques, and the amount of data on the quality of the learned representations. We also aim to use more downstream tasks and other techniques for assessing the quality of representations such as activation maximization and similarity analysis. Finally, we plan to investigate the potential of in-domain representation learning on various small datasets. For this, we will investigate the impact of the downstream setup, including the downstream model and the downstream data availability.

\bibliographystyle{IEEEbib}
\bibliography{strings,refs}

\end{document}